%% file: main.tex
\documentclass[conference]{IEEEtran}

\usepackage{tikz} 
\usepackage{listings}
\usepackage{enumitem}
\usepackage{graphicx}
\usepackage{multicol, latexsym}
\usepackage{blindtext}
\usepackage{caption}
\usepackage{longtable}
\usepackage{csquotes}
\usepackage{amsfonts}
\usepackage{amsmath}
\usepackage{amsthm}
\usepackage{amssymb}
\usepackage{algorithm}
\usepackage{tabularx}
\usepackage{capt-of,lipsum} 
\usepackage{balance}
\usepackage{subfigure} 
\usepackage{listings}
\usepackage{comment}
\usepackage{diagbox}
\usepackage{pdflscape}
\usepackage{booktabs}
\usepackage{makecell}

\usepackage[bookmarks=false]{hyperref}

\definecolor{myblue}{rgb}{0.09,0.20,0.34}
\definecolor{mygreen}{rgb}{0,0.6,0}
\definecolor{mygray}{rgb}{0.98,0.98,0.98}
\definecolor{myorange}{rgb}{0.92,0.49,0.34}
\definecolor{mywhite}{rgb}{1.0,1.0,1.0}

\definecolor{NMR}{RGB}{255,255,86}
\definecolor{MRA}{RGB}{255,231,27}
\definecolor{MRD}{RGB}{178,178,178}
\definecolor{MRIP}{RGB}{188,172,0}
\definecolor{MRS}{RGB}{161,207,106}
\definecolor{NA}{RGB}{228,60,52}
\definecolor{AI}{RGB}{255,255,86}
\definecolor{RMR}{RGB}{162,4,21}
\definecolor{RMD}{RGB}{178,178,178}
\definecolor{RMA}{RGB}{255,231,27}
\definecolor{RMIP}{RGB}{188,172,0}
\definecolor{RMS}{RGB}{161,207,106}
\definecolor{LIGHT_GREY}{RGB}{240,240,240}
\definecolor{lightgray}{rgb}{.9,.9,.9}
\definecolor{darkgray}{rgb}{.4,.4,.4}
\definecolor{purple}{rgb}{0.65, 0.12, 0.82}
\definecolor{darkgreen}{RGB}{30, 142, 20}

\newcommand{\ie}{\textit{i}.\textit{e}.,\ }
\newcommand{\eg}{\textit{e}.\textit{g}.,\ }
\newcommand{\cf}{\textit{c}\textit{f}.,\ }

\newcommand\MarkFirst{\textsuperscript{1}}
\newcommand\MarkSecond{\textsuperscript{2}}

\newcommand{\oracle}{\textit{WeTrace}}

\lstdefinelanguage{JavaScript}{
	keywords={typeof, new, true, false, catch, function, return, null, catch, switch, let, var, if, in, while, do, else, case, break, async, await, const},
	keywordstyle=\color{purple}\bfseries,
	ndkeywords={class, export, boolean, throw, implements, import, this},
	ndkeywordstyle=\color{darkgray}\bfseries,
	identifierstyle=\color{black},
	sensitive=false,
	comment=[l]{//},
	morecomment=[s]{/*}{*/},
	commentstyle=\color{darkgreen}\ttfamily,
	stringstyle=\color{darkgreen}\ttfamily,
	morestring=[b]',
	morestring=[b]"
}

\newsavebox{\mybox}

\usepackage{pifont}
\newcommand{\cmark}{\ding{51}}%
\newcommand{\xmark}{\ding{55}}%

\begin{document}




\title{\oracle \\A Privacy-preserving Mobile COVID-19\\Tracing Approach and Application}



\author{\IEEEauthorblockN{
        A. De Carli\MarkFirst, M. Franco\MarkSecond, A. Gassmann\MarkFirst, C. Killer\MarkSecond, B. Rodrigues\MarkSecond, E. Scheid\MarkSecond, D. Sch{\"o}nb{\"a}chler\MarkFirst, B. Stiller\MarkSecond}

\IEEEauthorblockA{
    \\
    \MarkFirst Papers AG, Dammstrasse 16,
    CH-6300 Zug, Switzerland \\
	\MarkSecond Communication Systems Group CSG, 	Department of Informatics IfI, University of Zurich UZH\\
	Binzm{\"u}hlestrasse 14, CH--8050 Z{\"u}rich, Switzerland\\
	\\
	E-mail: \{a.decarli,a.gassmann,d.schoenbaechler\}@papers.ch\\
	\{franco,killer,scheid,rodrigues,stiller\}@ifi.uzh.ch\\
	\\
	\today \\v2.0}
}


\newtheoremstyle{mydef}
{\topsep}{\topsep}%
{}{}%
{\bfseries}{}
{\newline}
{%
  \rule{\linewidth}{0.4pt}\\*%
  \thmname{#1}~\thmnumber{#2}\thmnote{\ -\ #3}.\\*[-1.5ex]%
  \rule{\linewidth}{0.4pt}}%
\theoremstyle{mydef}
\newtheorem{definition}{Definition}
\newtheorem{protocol}{Step}

            


\maketitle


\begin{abstract}
For the protection of people and society against harm and health threats --- especially for the COVID-19 pandemic --- a variety of different disciplines needs to be involved. The data collection of very basic and health-related data of individuals in today's highly mobile society does help to plan, protect, and identify next steps health authorities and governments can, shall, or need to plan for or even implement. Thus, every individual, every human, and every inhabitant of the world is the key player --- very different to many past crises'. And since the individual is involved --- all individuals --- his/her \textit{(a)} health and \textit{(b)} privacy shall be considered in a very carefully crafted balance, not overruling one aspect with another one or even prioritizing certain aspects. Privacy remains the key.

Thus, the solution of the current pandemic's data collection can be based on a \textit{fully privacy-preserving} application, which can be used by individuals on their mobile devices, such as smartphones, while maintaining at the same time their privacy. Additionally, respective data collected in such a fully distributed setting does help to confine the pandemic and can be achieved in a democratic and very open, but still and especially privacy-protecting world. 

Therefore, the \textit{WeTrace} approach and application as described in this paper utilizes the Bluetooth Low Energy (BTE) communication channel, many modern mobile devices offer, where asymmetric cryptography is being applied to allows for the deciphering of a message for that destination it had been intended for. Since literally every other potential participant only listens to random data, even a brute force attack will not succeed. WeTrace and its Open Source implementation is the only known approach so far, which ensures that any receiver of a message knows that this is for him/her, but does not know who the original sender was. 
\end{abstract}



%

\IEEEpeerreviewmaketitle

\pagestyle{plain} 


\input{sections/introduction}
\input{sections/relatedwork}
\input{sections/requirements}
\input{sections/architecture}
\input{sections/implementation}
\input{sections/observations}
\input{sections/conclusions}



\section*{Acknowledgements}
This paper was supported partially by \textit{(a)} the University of Z\"urich UZH, Switzerland and \textit{(b)} the European Union's Horizon 2020 Research and Innovation Program under Grant Agreement No. 830927, the CONCORDIA project. 




\balance
\bibliographystyle{IEEETranCustomized}
\bibliography{bib/references.bib}

\end{document}

%% file: sections/introduction.tex
\section{Introduction}

Social distancing is one of the essential measures to prevent the spreading of COVID-19 in a population. The current situation indicates that any step toward the re-establishment of the society’s regular economic activities shall be performed very carefully by health authorities and governments in order to prevent a second or new infection waves. Thus, the use of existing and already rolled-out technology is essential and almost the only way \textit{(a)} to crowd-source information concerning the health of individuals and \textit{(b)} to ensure that social distancing rules are being respected. These two requirements stem from the perspective of a general epidemic analysis and urgently need to be combined with technical support such that \textit{(i)} the individual’s privacy, \textit{(ii)} the personal freedom of an inhabitant of Switzerland and many other countries --- in contrast to a selected list of differently organized countries of the world ---, and \textit{(iii)} the avoidance of ``finger-pointing'' to select humans can be reached technically and efficiently in the shortest possible time frame.

For instance, the spreading rate of infection based on currently available data (\eg in how many days does the number of infected individuals doubles) may imply that as observed in some cases in Italy or Spain, it is not possible for individuals to be examined because the hospital's infrastructure may be overloaded. Thus, waiting for a doctor's signaling that an individual is infected may be too late to prevent further infections from that individual. Furthermore, the asymptomatic period, where individuals are not aware of his/her infections, can imply that a system may emit distance alerts regardless of whether a mobile device flagged an unidentified human being infected is nearby or not. 

As of today, most individuals carry such mobile, but different devices with different communication choices being integrated. These devices include phones –-- generally termed smartphones ---, tablets, and laptops. Each of them is connected through partially selectable communication technologies, such as Bluetooth (BT), Bluetooth Low Energy (BTE), Wi-Fi (IEEE 802.11 family of protocols), Wi-Fi direct, GSM (Global System for Mobile Communications), UMTS (Universal Mobile Telecommunication System), HSDPA (High-Speed Downlink Packet Access), LTE (Long-term Evolution), or LTE-A (Long-term Evolution Advanced) (often termed 3G, 4G, and partially already moving into 5G technologies). Many of those alternatives allow for the communication of a mobile device's position to nearby devices and even further on. 

For this work on WeTrace, the fully privacy-preserving approach and application, the focus is on BTE, since all 3G to 5G communication technologies deployed already will allow for an identity tracking of this device used, since the SIM (Subscriber Identity Module) card-based identification of communications can always reveal the current user’s (subscriber) identity, who was registered with the respective contract. Thus, a clear demand beyond public telecommunication system-based communications for a full privacy-preserving tracking and tracing is very urgently required. Note that WeTrace potentially offers as well a solution to meet major GDPR (General Data Protection Regulation) requirements, which are in force in certain European countries. 

The WeTrace approach and application developed here fulfills exactly this key requirement on privacy-preserving for arbitrary mobile devices, being able to communicate via BTE and being used by their owners in a once-used, once-associated manner. This means that the underlying assumption of this work is --- as very many others do that, too, but do not necessarily state that explicitly: ``one mobile device belongs to one individual and is used by this individual only during the pandemic''. Furthermore, the application of low-range BTE communications determines a highly suitable coincidence between the COVID-19 ``social distancing'' requirements and the communications technology: Since only those individuals in a range of a few meters (if staying that close together for approximately 10 to 15 minutes) potentially are subject to contagious infection, their mobile devices can, if the WeTrace application is enabled and configured, exchange health data in a fully privacy-protecting manner, such that the infection status information can be exchanged fully anonymous and in a secured (encrypted) manner. 

Finally, WeTrace as a tracing approach and application will not deprecate over a very short period of time, since the fully privacy-preserving characteristic can serve for future health-related as well as other data collection applications, such as high-density events in entertainment cases or natural disasters like earthquakes or tsunamis, where many independent individuals can report and a respective view on locally residing individuals can be fed back. 

\vspace{\baselineskip}


This paper is organized as follows. While Section~\ref{sec:related_work} summarizes major related work and compares those against the key dimensions of utmost importance set for a privacy-preserving tracing application, Section~\ref{sec:requirements} determines the essential requirements needed to meet the goal of hiding in full the privacy of an individual participating. Derived from those requirements the WeTrace approach is introduced in Section~\ref{sec:stakeholders_architecture} and relevant technical details are provided. Section~\ref{sec:implementation} follows with major specificities of its technical implementation. While  Section~\ref{sec:observations} discusses first trade-offs taken as well as observations obtained and analysis performed with respect to proximity, privacy, and user overhead, it complements second with evaluations on the communication channel used, attacks mitigated, storage sizes, and scalability. Thus finally, Section~\ref{sec:conclusions} summarizes and concludes the WeTrace work as one essential step into a fully privacy-preserving tracking and tracing application world.


%% file: sections/relatedwork.tex
\section{Related Work}
\label{sec:related_work}

Since the formal declaration of COVID-19 being a worldwide pandemic, progressing at a very fast pace, a set of different projects and tools have been proposed to implement automated notification or alerting systems as a supporting tool to ensure the re-establishment of society’s normal activities, while safely maintaining the human's health as the primary goal. Of course, systems for tracking and tracing exist and various mobile devices show a variety of very different, but related functionality, including the search option for a displaced smartphone, once it had been misplaced or stolen. However, very many if not all such approaches known do trade-off the individual's privacy or the human's data privacy in one way or another. Thus, the notification or alerting option of today's tracking and tracing systems has not reached an acceptable standard with respect to user privacy and user data privacy. 

Thus, it is highly important, especially with respect to an application collecting health data, to observe security and more precisely privacy concerns of all solutions proposed so far, which address COVID-19 tracing anonymously. Privacy includes the privacy of the user and the privacy of user data, here a medical status. Additionally, this includes for sure non-negotiable guarantees that these systems cannot be used as \textit{(a)} attacking tools to the availability of other systems, \textit{(b)} jeopardizing the privacy of users themselves, \textit{(c)} violating the privacy of user data, and \textit{(d)} revealing at any step performed private and to be secured personal (health) data. In this sense, Table~\ref{tab:related_work} collects the major characteristics of related systems in order to clarify the state-of-the-art concerning tools and proposals related to COVID-19-driven tracing and tracking. An extensive report of further related work collections performed by several contributors can be found at~\cite{googleDocsRw}. A global report on approaches by governmental and private projects to use personal data to combat COVID-19 can be found at~\cite{GDPRhubRw}. However, since this work here on WeTrace addresses the user privacy and user data privacy as its first priority, the constructive work to define and specify an appropriate solution is considered to be more relevant at this stage than collecting yet another complete view of country-specific approaches under way these days. 

Nevertheless, since the pandemic is very recent and is spreading in some countries of the world at a very fast pace (an exponential growth was observed in certain countries, in many others is had slowed down slowly at the time of writing~\cite{COVID-19-Dashboard}), it is important to note that most of these approaches as referred to above are theoretical proposals or are still under development. Thus, the detailed information on these technological solutions and approaches might become deprecated over a very short period. Clearly, the use of more general and generic applications will help to sustain the work and research invested. As such, the Sismo approach~\cite{sismo} was proposed for earthquake notifications and now is being used to operate for COVID-19 notifications. 

The key dimensions of the table on related work compiled here and their detailed description are determined in different dimensions as follows:
\begin{enumerate}
    \item \textbf{Solution}: Proposed tool or respective solution, including its current name(s) and currently available key reference(s).
    \item \textbf{Open-Source}: Determination whether the code is or will become publicly available. This is essential for a verification of the privacy-preserving property of the approach as well as other security metrics
    \item \textbf{Reporting}: Concerning whether users applying this approach within their smartphone are able to flag themselves being in one certain medical state, \eg ``not infected'', ``close contact with infected'', ``infected'', ``infected, being with symptoms'', ``indifferent'', or ``healthy''. There is no ``cured'' state, as this state is considered as ``not infected''. However, since the actual states are not important for the WeTrace application's operation (any sort of data can be integrated into a message,  unless it grows to large for the Backend to publish), the actual medical states needed have to be discussed and defined with epidemiologists. 
    \item \textbf{Data Collected}: Determining the type of data being collected and processed, generally that is possible from the user and/or a Backend (cf. below), if involved. These data may include Global Positioning System (GPS) data for geo-localization, timing-related information, medical status (cf. before), communication addresses, or phone numbers.
    \item \textbf{Privacy-preserving Mechanism}: As it is mandatory that any solution/tool does not violate the users’ privacy, \eg by revealing their identity, health conditions, or geographical localization. A security and risk analysis of this or these mechanism(s) foreseen or deployed does determine the level of privacy to be reached. 
    \item \textbf{Communication Technology}: The technical communication solution selected enables the collection of relevant (or irrelevant) information on the users’ encounters. Mainly the focus here is on BLE and GPS. Approaches using of the Wi-Fi communication technology or any 3G to 5G communication approach is not listed here, since all of them do limit by definition the users’ privacy due to the use of IP (Internet Protocol) addresses or phone numbers, since Internet Service Providers are required by law in selected countries to keep track of the ``owner'' of an IP address for a certain time and the use of SIM cards legally requires a registration of users with their full identity, respectively.  
    \item \textbf{User Notification}: The key feedback channel back to the user needs to be identified, especially for any type of feedback the user may want to know about the data collected, \eg graphs, statistics, summaries, or simple ``encounter'' information. Depending on the system and data, such data may be already privacy-protected, thus, encrypted. 
    \item \textbf{Storage}: The storage of Data Collected is partially important for comparisons, statistics, and trends. Thus, different approaches are deployed to store data in general or messages specifically in form of: \eg local storage, in the cloud, on private servers, or on individual devices only. Also, based on the storage approach, Backends (cf. next) might be required to receive and actively forward information related to the medical status.
    \item \textbf{Backend}: The Backend is important for \textit{(a)} an exchange of data between devices, which are geographically not close (any more due to mobile users), \textit{(b)} a possible ``comparison'' of data broadcast to the Backend, \textit{(c)} a pure relaying of messages, or \textit{(d)} a publication of static content. Depending on the specific role intended, one can derive how much power and information the provider of such a Backend holds. Example instances of these (not necessarily orthogonal) roles include servers with \eg a central database of all medical states, a relay functionality for messages, a broadcasting function for messages, or a publishing activity of static content.
\end{enumerate}


%% file: sections/requirements.tex
\section{Requirements}
\label{sec:requirements}

Technical requirements of a tracking and tracing application are explicitly extracted based on COVID-19 pandemic characteristics and are outlined to determine \textit{(a)} the minimal set of functionality needed and \textit{(b)} major system boundaries and constraints of the WeTrace application, which are defined in terms of user privacy. The respective design following in later sections of the paper complies to such requirements. Especially system boundaries and constraints taken into consideration need to comply to GDPR regulations and the respective user data and user privacy. 

Despite the main concern with the users' confidentiality, as well as the legal aspects to which a mobile application is submitted, other requirements such as usability, scalability, and energy efficiency, are also relevant to determine its success in terms of mass adoption. Thus, simplicity is the key to make intuitive user interfaces and to avoid unnecessary operations (\eg intensive and explicit use of BLE or GPS). 

Besides essential requirements for such an application and its implemented system of being epidemiological sensible and useful, important soft-requirements exist. Especially in cases where people utilize a tracing application on a voluntary basis, the importance of those soft-requirements becomes clear, since volunteers installing such an application need to be ensured that such an application does comply to all of the following ones. Thus, the following list of requirements was identified as crucial:

\begin{itemize}
    \item Privacy
    \item Scalability
    \item Energy Consumption
    \item User Overhead
    \item Legal Compliance
\end{itemize}

\input{tables/related_work}

\subsection{Privacy Properties}
\label{sec:privacy_properties}

Moreover, within the general context of privacy, the properties defined by~\cite{cho2020contact} are taken into consideration for the design of the WeTrace application, too, which explicitly include:

\begin{enumerate}
    \item Privacy from Snoopers
    \item Privacy from Contacts
    \item Privacy from Authorities
\end{enumerate}

These three properties determine three dedicated and potential attack vectors, since any of these three roles listed potentially could harm the system's coherent and trustworthy operation. Thus, they are evaluated in Subsection~\ref{ssec:privacy_enforcements}.

\subsection{Scalability}

Within a pandemic setup it is expected that data, \ie of new positive cases, can grow exponentially in a very short amount of time. WeTrace needs to be able to cope with this exponential growth in order to be useful, when it is needed the most. Thus, \textit{(a)} the number of infections, \textit{(b)} the number of ``close contacts'', and \textit{(c)} the number of keys determine relevant parameters impacting WeTrace's scalability. A fourth scalability dimension is determined in terms of regions covered. Without any doubt, achieving an application suitable also across multiple countries will be inherently more useful. A selected set of numerical examples related to the scalability of WeTrace is discussed in Section~\ref{sec:evaluations}.

\subsection{Energy Consumption}

Even though the scanning and advertising BLE packages has a minimal impact on a smartphone's battery life, compared to communication alternatives such as ZigBee/802.15.4~\cite{bleZigBee}), it is evident that a user that allows the WeTrace application to run in the background will not perceive the application as acceptable, if it is draining the battery life of his/her device. In this sense, it is imperative to be compatible with significant battery optimization mechanisms available on the Android and iOS platforms. Hence, the WeTrace implementation has to consider the impact on battery life with the use of BLE for tracking close contact encounters \cite{impactBLEBattery,advPowerConsumption}.

\subsection{User Overhead}

The user will have to receive the WeTrace application from a trusted platform, which means a minimal overhead for all users to launch and leave it running in the background. Ideally, the WeTrace protocol can be included into already accepted and existing apps, so that the user does not need to install any additional application to ensure his/her privacy. However, if the WeTrace application is installed separately, the installation experience needs to be very simple for any user to get started easily. An approach of such a WeTrace application integration related to the user overhead minimization is discussed in Section~\ref{sec:evaluations}.

\subsection{Legal Compliance}

The legal compliance with data protection laws and regulations, \eg the GDPR, is crucial for any technical solution that collects and analyzes user data \cite{DP3T}. Specifically, Article~25 of the European Union (EU) GDPR states that \textit{only personal data, which are necessary for each specific purpose of the processing, are processed. That obligation applies to the amount of personal data collected, the extent of their processing, the period of their storage, and their accessibility} \cite{gdprart25}. Thus, any digital tracing solution must respect the legal requirements, minimizing the data collected and processed. In case of WeTrace this goal is reached, since as designed and documented below, it is the user who decides on data to be shared besides the fact of a ``close contact'' in a fully anonymous manner. 

According to Article 4(15) of the EU GDPR \cite{gdparart4}, \textit{data concerning health} concerns any personal data related to the physical or mental health of a natural person, thus, this also applies to WeTrace. Since, the WeTrace messages are encrypted with the public key of the Device $B$, WeTrace assures the confidentiality of all messages communicated (thus. published) via the Backend (\cf \ref{fig:sequence_diagram}). Hence, negligent parties (\eg Device $C$), cannot decrypt data seen in this message received, since these messages are encrypted with the public key received via the ``close contact'' over a BLE advertisement packet. 


%% file: tables/related_work.tex
\begin{landscape}
\vfill
\begin{center}
\begin{table}[htb]
\caption{Tracing Approach and Application Comparisons}
\label{tab:related_work}
\begin{tabular}{c|c|c|c|c|c|c|c}
\toprule
\multicolumn{1}{c|}{\textbf{Solution}} & 
\multicolumn{1}{c|}{\makecell{\textbf{Open}\\\textbf{Source}}} & 
\multicolumn{1}{c|}{\textbf{Reporting}} & 
\multicolumn{1}{c|}{\textbf{\centering Data Collected}} &
\multicolumn{1}{c|}{\makecell{\textbf{Privacy-preserving}\\ \textbf{Mechanism Applied}}} & 
\multicolumn{1}{c|}{\makecell{\textbf{Communications}\\ \textbf{Technology}}} & 
\multicolumn{1}{c|}{\centering \textbf{Data Storage}} &
\multicolumn{1}{c}{\centering \textbf{Backend}} 

\tabularnewline

\toprule
\hline

WeTrace	& Yes & \centering Self-reporting & \makecell{GPS location history, \\ encounter timestamp} & \makecell{Public Key Cryptography,\\ GDPR-compliant privacy, \\ fully anonymous} & BLE &	\makecell{Decentralized, \\ locally at the device} & \makecell{Run by either authorities \\ or trustworthy institutions, \\ broadcasting of notifications} 
\tabularnewline
\hline

CoroTrac~\cite{CoroTrac} & No & \centering Self-reporting & \makecell{GPS location history} & \makecell{Data anonymization model} & GPS & \makecell{Centralized, \\ own database} & \makecell{Infrastructure maintained by \\ the developer's institution}
\tabularnewline

\hline

CovidWatch~\cite{CovidWatch} & Yes & \centering Self-reporting & \makecell{GPS location history} & \makecell{GPS anonymization model} & BLE & \makecell{Decentralized, \\ locally at the device} & \makecell{Public database maintained by \\ the developer's institution, \\ broadcasting of notifications} \tabularnewline

 \hline

Pandoa~\cite{Pandoa} & Yes & \centering Self-reporting & \makecell{GPS location history} & \makecell{GPS anonymization model} & BLE, GPS & \makecell{Centralized, \\ own database} &  \makecell{Infrastructure maintained by \\ the developers, send notifications \\ for possible contacts}
\tabularnewline

\hline

NextTrace~\cite{NextTrace} & No & \makecell{Provided by Labs \\ and self-reporting} & \makecell{Location and \\ proximity data} & \makecell{Data anonymization model} & \makecell{BLE, GPS}  & N/A & \makecell{Infrastructure maintained by \\ the developers, send notifications \\ for possible contacts}

\tabularnewline

\hline

geoHealthApp~\cite{geoHealthApp} & No & \makecell{Claimed to be \\ AI-based} & \makecell{GPS location history} & \makecell{Blockchain, claimed GDPR \\ compliant} & GPS & \makecell{Centralized, \\ own database} & \makecell{Infrastructure maintained by \\ the developers} \tabularnewline

\hline

CoronaTrace~\cite{CoronaTrace} & No & \centering Self-reporting & \makecell{GPS location history} & \makecell{User data anonymization, \\ not publicly visible \\ individual information} & GPS & \makecell{Centralized, \\ own database} & \makecell{Infrastructure maintained by \\ the developers, send notifications \\ for possible contacts}
\tabularnewline

\hline

TraceTogether~\cite{TraceTogether} & No & \centering Self-reporting & \makecell{Location and location \\ of nearby devices} & \makecell{\centering Encrypted BLE channels} & BLE & \makecell{Decentralized, \\ locally at the device} & \makecell{Infrastructure maintained by \\ the developers, send notifications \\ for possible contacts}
\tabularnewline
\hline

\makecell{DP3T~\cite{DP3T} \\ part of PEPP-PT~\cite{pepp}} & Yes & \centering Self-reporting & \makecell{GPS location history, \\  encounter timestamp} & \makecell{Apply Hash functions, \\ pseudoanonymous for \\ research volunteers, GDPR} & \makecell{GPS, Cell phone \\ triangulation , BLE} & \makecell{Decentralized, \\ locally at the device} & \makecell{Run by either authorities \\ or trustworthy institutions, \\ broadcasting of notifications} \tabularnewline
\hline

\makecell{NextStep~\cite{NextStep} \\ part of PEPP-PT~\cite{pepp}} & Partially & \centering Self-reporting & \makecell{Only P2P Bluetooth \\ encounters} & Encrypted (not disclosed) & BLE & \makecell{Decentralized,\\ locally at the device} & \makecell{Full details not disclosed, matching \\ of IDs all done on device}
\tabularnewline
\hline

NOVID20~\cite{novid20} & Yes & \centering Self-reporting & \makecell{GPS location history, \\  Bluetooth, and Google} & \makecell{Encrypted (not disclosed)} & BLE, GPS & \makecell{Decentralized, \\ locally at the device} & \makecell{Infrastructure maintained by \\ the developers, send notifications \\ for possible contacts}
\tabularnewline

\hline

StopCorona~\cite{stopcorona} & No & Self-Reporting & \makecell{Mobile number for 30 days,\\ Bluetooth, nearby audio} &  \makecell{Pseudonymous, collects \\ user information for \\  research purposes} & Microphone, BLE & Centralized & \makecell{Infrastructure maintained by \\ the developers, notifications \\ available publicly}
\tabularnewline

\hline

Sismo~\cite{sismo} & No & \centering Self-reporting & \makecell{GPS current location} &  \makecell{Pseudonymous, collects \\ user information for \\ research purposes} & GPS & \makecell{Centralized,  \\ cloud-server} & \makecell{Infrastructure maintained by \\ the developers institution} \\

\bottomrule
\end{tabular}
\begin{footnotesize}
\end{footnotesize}
\end{table}
\end{center}
\vfill

\end{landscape}


%% file: sections/architecture.tex
\section{WeTrace Stakeholders and Architecture}
\label{sec:stakeholders_architecture}

Taking the requirements listed in Section~\ref{sec:requirements} above into consideration, this section presents the WeTrace application to address the problem of privacy in tracking of COVID-19 cases. WeTrace is a fully privacy-preserving solution that relies on cryptographic mechanisms, such as asymmetric key pairs, to provide an application, where identities of users are only known by the user him/herself. Thus, this section details the stakeholders involved, presents the WeTrace architecture, and describes the flow of information between all major components and stakeholders.

\subsection{Stakeholder Definitions}
\label{ssec:stakeholder_definitions}

Even though WeTrace consists of a simple approach involving \textit{Users}, \textit{Devices}, and a \textit{Backend}, there are other relevant and related stakeholders to be considered. These, including the main three ones, are described as follows:

\begin{itemize}
    \item \textbf{Users} are individuals using the WeTrace application, for which any person can have at this state of th implementation three possible states: \textit{(i)} ``not infected'', \textit{(ii)} ``close contact with infected'', or \textit{(iii)} ``infected''. As discussed above in Section~\ref{sec:related_work}, the actual states are to be discussed with epidemiologists.
    \item{\textbf{Medical Doctor}}: Currently, the WeTrace application relies on self-reporting to detect COVID-19 infections, which might lead to the spam of false-positives. To address such a problem, a medical doctor does act as a testing person. However, this could potentially weaken the privacy of the approach, if doctor-user relations may become public. 
    \item{\textbf{Governmental Health Agency}}: Similarly to medical doctors, governmental health agencies could provide trusted data concerning the medical status of an individual. However, research on how to maintain privacy concerning eHealth data must be conducted, as pointed out by~\cite{healthPrivacy,healhAnonymization}. Thus, the WeTrace application focuses at this stage of the implementation on those three medical states as determined above only. 
    \item \textbf{Devices} determine the technical platform on which GPS and BLE-enabled communications happen and where WeTrace is installed on. A Device stores the following information: Master seed used to generate public-private key pairs, public keys of devices (which have WeTrace installed) encountered within 2~m of proximity and being in contact for longer than an epidemiological relevant time (\eg 10 to 15~min), a timestamp, and the approximated geo-location of encounters.
    \item A \textbf{Backend} broadcasts all messages of users who changed their status from ``not infected'' to ``infected''. The Backend does not store any data, only publishes them to other WeTrace-enabled devices, which perform the decryption of messages in case of needs.
    \item The \textbf{Server Administrator} of the Backend must be taken into consideration as he/she will have access to all messages originating from WeTrace applications. Although all messages are encrypted with private keys and the server does not store any public keys, the server administrator has full access of these messages and related information, \eg IP addresses and the device platform that could be used to infer the identity of a user. 
    \item \textbf{Identity Provider:} Even though the identity of the user is only known within the device (based on the storage of private keys generated via the master seed), such devices require outside the use of the WeTrace application the registration with \textit{Communication Service Providers}, \eg telephony operators, 4G cells and their Base Stations, or \textit{Internet Service Providers (ISP)} with their Wi-Fi networks. Thus, the identity of the device a user utilizes could be retrieved, if such providers share this information for a meta analysis.
\end{itemize}

\begin{figure*}[!ht]
    \centering
    \includegraphics[width=0.9\textwidth,keepaspectratio]{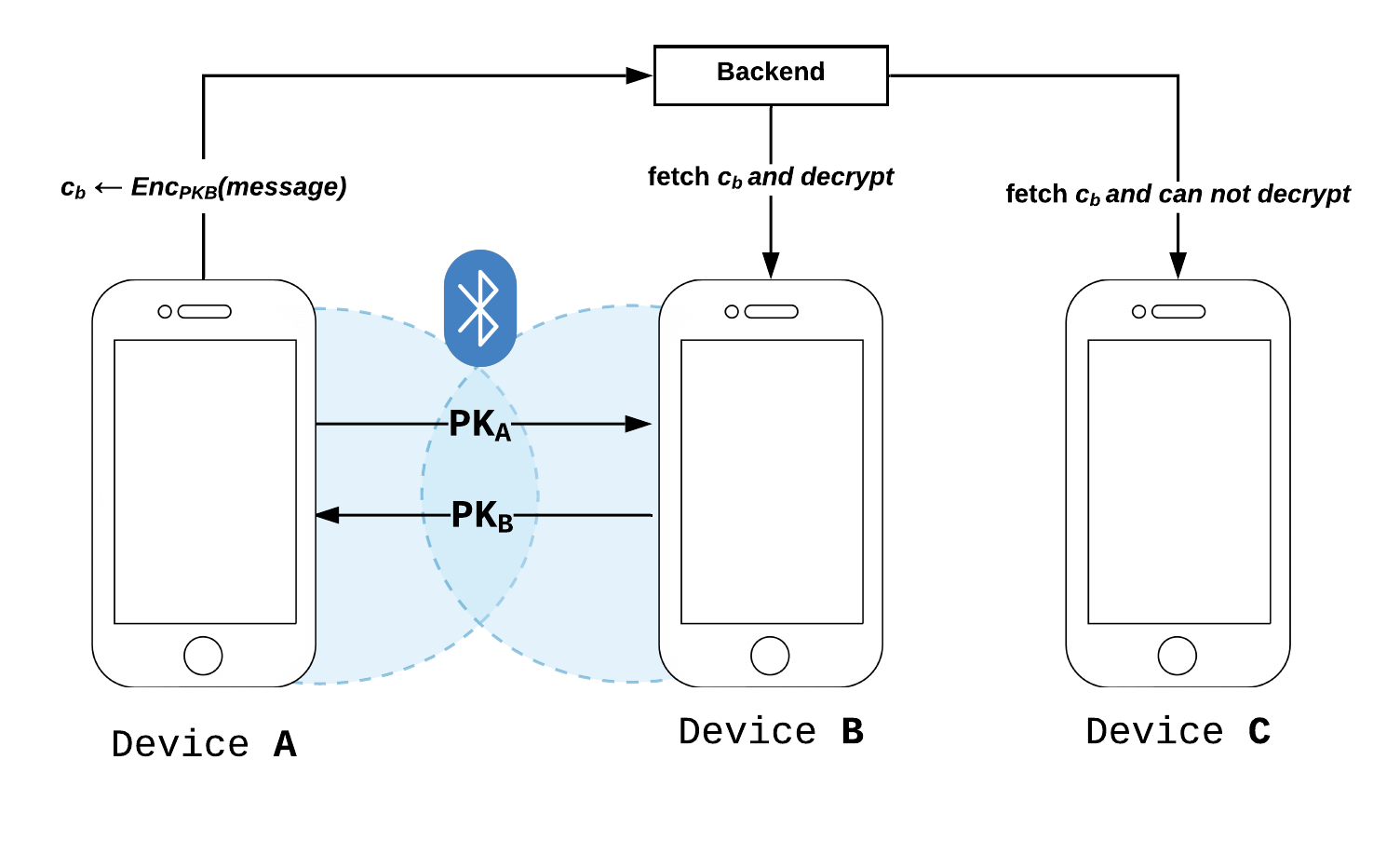}
    \caption{WeTrace Architecture}
    \label{fig:arch}
\end{figure*}

\subsection{Architecture}

Figure~\ref{fig:arch} depicts the architecture of WeTrace as proposed. Three devices are shown, Device A, Device B, and Device C. All three devices have the WeTrace application installed and are broadcasting their public key. 

Whenever these devices see another device that is broadcasting its public key over BLE, they will store that public key received in their local storage. Furthermore, all devices actively poll the latest encrypted messages from the Backend and try to decrypt them. This connection to the Backend happens through a general Internet connection, since only the exchange of public keys happens through BLE. Finally, encrypted messages are published by all Devices via the Backend, which will then make these newly encrypted messages available for everyone else, thus, all other devices. The third party hosting the Backend potentially will be able to collect IP addresses from those communications and devices, which are publishing encrypted messages. Even though the content of those messages cannot be read by the third party the fact that they can be linked to IP addresses can potentially pose a threat to privacy. To avoid this from happening on that level, a possible mitigation routes the publishing of messages from a device through the Tor network.

\subsection{Sequence Diagram}

\begin{figure}
    \centering
    \includegraphics[width=1\linewidth,keepaspectratio]{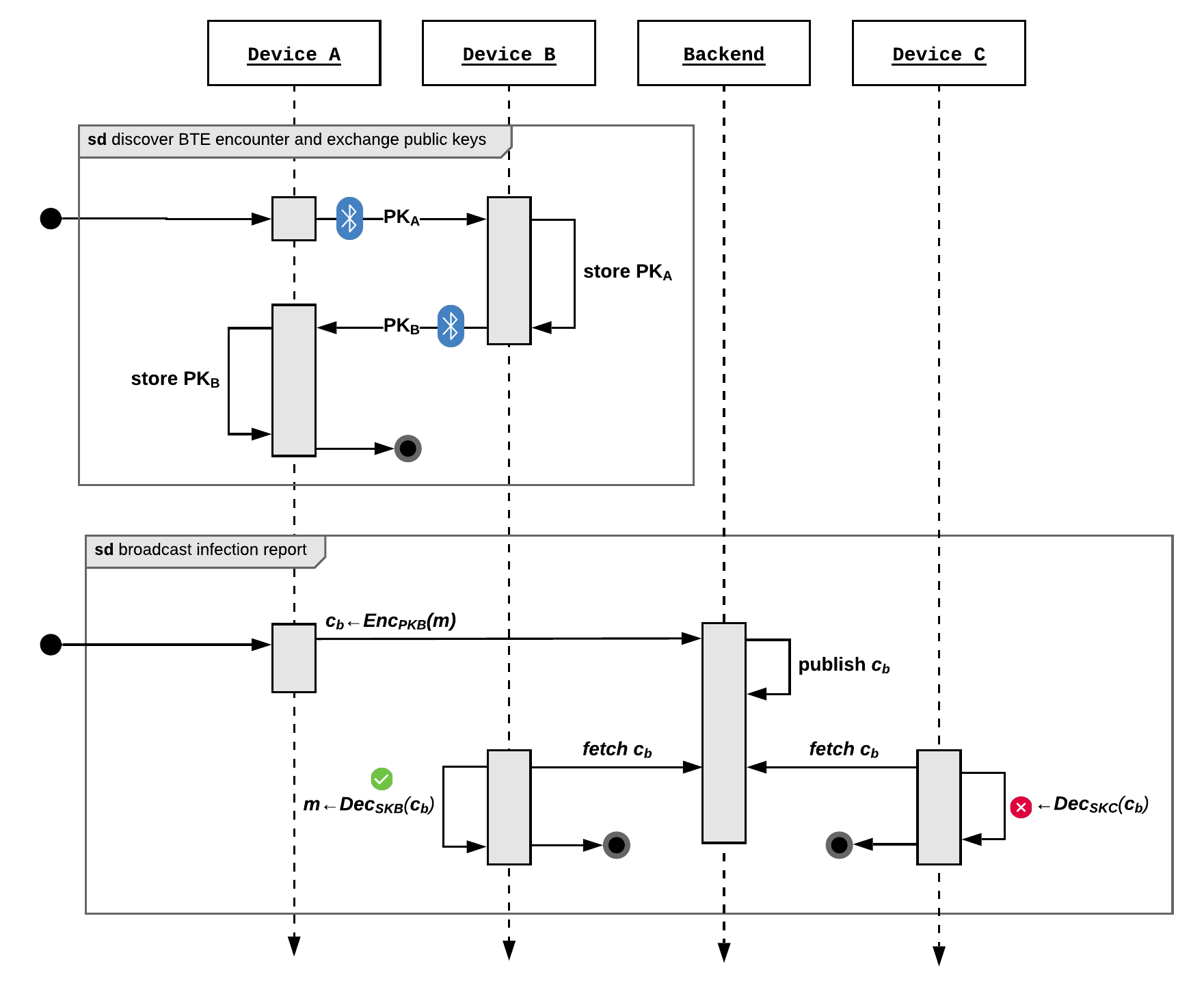}
    \caption{WeTrace Sequence Diagrams}
    \label{fig:sequence_diagram}
\end{figure}

The overall flow of information and interaction between the main stakeholders of WeTrace is shown in Figure~\ref{fig:sequence_diagram}. On one hand, Device A and Device B  ``see each other'' for a long enough time, such that they decide independently from each other to store the other party's public key in their local storage on the device. On the other hand, Device C is not in reach of those Devices A and B, hence Device C neither collects public keys nor advertises the own public key to anyone successfully. In case a fourth Device D may pop up in proximity to Device C, the public key advertised by Device C regularly will be stored by Device D and vice versa.  

Under the assumption that Device A decides to report an infection, it will iterate locally through all the public keys stored within the local storage and will encrypt the new message containing the information the user decides to broadcast with each public key individually, such as at least the infection status. Among those public keys locally stored there will also be the public key advertised by Device B previously.

Since Device B will poll from the Backend regularly the latest messages, it will try to decrypt those ones received. Thus, Device B will then find out that one message can be decrypted with his/her own private key, meaning that the message was intended for this Device B only. At that point in time the WeTrace application has to inform the user of Device B that the message was intended for this device and contains the decrypted information as communicated.

In the meantime, Device C will also poll messages from the Backend regularly, however, since the his/her public key was never collected by Device A, none of these messages received can be decrypted. Hence, Device C will know that no message was intended for it.

%% file: sections/implementation.tex
\section{Technical Implementation}
\label{sec:implementation}

WeTrace is proposed to be operational as an application on  mobile devices, typically a smartphone on iOS or Android basis. WeTrace deploys mechanisms and technology, which exists today and has proven to work in many instances, thus, no newly developed technology will be required. All stakeholders as defined in Subsection~\ref{ssec:stakeholder_definitions} above are considered to be operational and in existence. The open-source WeTrace implementation is available at~\cite{WeTrace} and serves as the basis all aspects described in the following. 

\subsection{WeTrace Properties}
\label{ssec:wetrace_properties}

Before WeTrace's technical assumptions are made explicit in Subsection~\ref{ssec:underlying_technical_assumptions} below and the respective security and attack models are outlined in Subsection~\ref{ssec:security_and_attack_model} below, the major WeTrace application properties are defined as follows: 

\begin{itemize}
    \item All data of ``close contacts'' stay only on the mobile device, these are never shared with a Backend.
    \item Locally collected ``close contacts'' (\textless2~m) are stored together with the timestamp of when the contact had happened and an approximate geolocation of where the contact had happened.
    \item In case of an ``infection report'' being received from a third party, \eg a doctor or a hospital with which the owner of the mobile devices had medical interrogations with, only those users, who had been in close contact will be notified (cf. Figure~\ref{fig:sequence_diagram}). The Backend sees those messages being encrypted, cannot read its content, and operates with the sole purpose of relaying the message.
    \item The user, who received a report on an infection or a non-infection, can decide whether he/she wants to:
    \begin{enumerate}
        \item Report only the infection to the close contacts of the last 14 days. This value of a constant of the protocol will be configured within the application upon deployment and defines a medical consensus defined by epidemiology.
        \item Report the above \textit{and} the timestamp of when that close contact had taken place.
        \item Report the above \textit{and} the geolocation of where the close contact had taken place.
        \item Report the above \textit{and} the timestamp of when that close contact and the geolocation of where the close contact had taken place.
    \end{enumerate}
\end{itemize}

By utilizing a suitable combination of BLE protocol features --- available in many mobile devices --- and security mechanisms --- to be made available via the WeTrace application --- in addition, these concepts in conjunction define the underlying system's architecture.

\subsection{Underlying Technical Assumptions}
\label{ssec:underlying_technical_assumptions}

The device in used for the WeTrace application is assumed to be trusted, since the secret key (used as the master seed to derive multiple public keys) is managed by the Operating System (OS) (\eg Android or iOS) and stored at the Secure Enclaves provided by the mobile OS. Furthermore, the general assumptions applying to asymmetric cryptography also apply to WeTrace, since public-key cryptosystems rely on the hardness of factoring discrete logarithms \cite{smart16}. Additionally, trust in the integrity of the GPS location and timestamp is also assumed, since it is generated by the underlying OS, using respective sensors of the smartphone. The underlying communication protocol BLE and especially the BLE Advertising Packets \cite{BLEAdvertizing} are deployed for exchanging close encounters are applied. Considering the WeTrace architectural depicted in Figure~\ref{fig:arch}, the following assumptions are listed:

\begin{itemize}
    \item The \textbf{Device} is assumed to be trusted, because secret keys must remain private, and, thus, data generated and broadcasted is trustworthy.
    \item The \textbf{Backend} is assumed not to be trusted concerning confidentiality, integrity, and availability. Therefore, devices are required to encrypt any message, including notification messages, with the public key of devices that had been received in close proximity over a defined period of time. Therefore, a interaction scheme is applied, where public keys are exchanged regularly, while the devices are near and reachable via BLE.
    \item The \textbf{Communication Channel} is provided by BLE's advertisements, which are assumed not to be trusted. Thus, a service provider cannot assure the availability of the service, \ie messages can fail to be delivered, and their content is readable from anyone receiving the message if not an explicitly encrypted payload is maintained. While this reflects the communication channel among devices BLE, the Device-to-Backend communication channel uses the Internet and is assumed to be not trusted as well. In the same way message content is readable from anyone receiving the message if not an explicitly encrypted payload is maintained. Furthermore, revealing of an identify of that device running this communications may be possible with external meta-data from the service provider, such as IP addresses or phone numbers. This is mitigated by applying the sending of messages to the backend via the tor network. 
\end{itemize}

\subsection{WeTrace Application Operation --- Example}

The WeTrace application's operation is exemplified within a given use case. Thus, for this example User A with Device A, User B with Device B, and User C with Device C are considered.

\begin{enumerate}
    \item Every device that installs the WeTrace app generates an asymmetric key pair using elliptic curve cryptography. For this examples $ PK_A $ stands for the public key of Device A and $ SK_A $ stands for the secret key of Device A. Thus, this step generated $ PK_A $, $ PK_B $, $ PK_C $, and $ SK_A $, $ SK_B $, $ SK_C $, respectively.
    \item In turn, every device starts broadcasting their $PK_*$, which is also considered to be their unique identifier to its surrounding devices.
    \item Upon devices (\ie A and B) now getting into ``close contact'' with User A, the Device A get to know $PK_B$ and the Device B gets to know $PK_A$ via the exchange of these over the BLE protocol. Besides these pubic keys $PK_*$, both devices also store a timestamp and the approximate geolocation of where this encounter had happened. Generally, this contact information is collected for all other devices in close proximity, too. 
    \item Upon User A's information received, \eg from a doctor or a hospital with which the owner of the device had medical interrogations with, User A wants to report an ``infection'' or a ``non-infected'' status. 
    Thus, the Device A will go through the list of close contacts collected as within step 3 and encrypts one message each for every public key of close contacts. In case of User A, it will be encrypted once with $PK_B$, because this was the only contact. 
    All those messages will be sent to the Backend via a different than BLE communication channel. The Backend will relay these messages to all devices. 
    The messages will contain the data that User A chose to share, so either only the fact that an infection happened or not, or additionally when or even where it happened. As noted above, it is important that only the reporting user decides if he/she wants to share this additional information.
    \item Device B and Device C now receive from the Backend a notification telling them that new reports have been seen. Thus, Device B will then try to decrypt every message already recorded at the step 3, the close encounter, with $SK_B$ and will eventually find out that a message was directed at him/her, thus, indicating to User B the status User A now knows and reported. 
    Device C will do perform exactly the same steps, however, because no message in his/her local storage was encrypted with $PK_C$, no message can be decrypted, thus, no information is relayed to User C.
\end{enumerate}

While this example indicates clearly how the use of public-private cryptography mechanisms ensures the generation of an unrecorded --- thus, no individual identity being assigned to a public-private key pair --- identity, which is temporary for the life-time of the application. A new installation will generate a new public-private key pair. The example shows as well that the Backend does not maintain the content nor any data, even if it would do so, run as a possible attack, the content will remain encrypted and unusable, since no proximity information would be available to that server. 

\subsection{Security and Attack Model}
\label{ssec:security_and_attack_model}

The WeTrace design does consider a wider spectrum of potential vulnerabilities and includes respective countermeasures. Therefore, to provide the technical means necessary to ensure user privacy and user data privacy, it is mandatory \textit{(a)} to ensure that at least one Backend is available to broadcast notifications to others and \textit{(b)} to ensure the integrity of these notifications sent by Devices with a minimal Backend interaction. Furthermore, the WeTrace Backend (as outlined above in Figure~\ref{fig:arch}) is assumed not to be trusted as such. Therefore, with respect to confidentiality, integrity, and availability the Table~\ref{tab:cia-analysis} discusses those characteristics for WeTrace Devices and the Backend. 

\begin{table}[]
\caption{CIA Analysis of the Backend and Devices}%
\resizebox{\columnwidth}{!}{
\begin{tabular}{c|l|l|l|l}

\toprule

\textbf{CIA}             & \multicolumn{2}{c|}{\textbf{Backend}}                                                                        & \multicolumn{2}{c}{\textbf{Devices}}                                                                                                   \\  \toprule \hline
\textbf{Confidentiality} & \begin{tabular}[c]{@{}l@{}}Does not provide  \\ guarantees\end{tabular}                                  & \xmark & \begin{tabular}[c]{@{}l@{}}Notifications are encrypted\\ with PK of previously in \\ range devices\end{tabular}                & \cmark \\ \hline
\textbf{Integrity}       & \begin{tabular}[c]{@{}l@{}}Does not provide\\ guarantees on the delivery\\ of notifications\end{tabular} & \xmark & \begin{tabular}[c]{@{}l@{}}Does not provide\\ integrity guarantees\\ of notifications\end{tabular}                             & \xmark \\ \hline
\textbf{Availability}    & \begin{tabular}[c]{@{}l@{}}Backend can fail or \\ be a target of DDoS \\ attacks\end{tabular}       & \xmark & \begin{tabular}[c]{@{}l@{}}Not ensured that neither \\ the sending or \\ receiving devices would\\  be available.\end{tabular} & \xmark \\

\bottomrule
\end{tabular}
}
\label{tab:cia-analysis}
\end{table}

\vspace{\baselineskip}
\subsubsection{Adversary Model}
WeTrace assumes an adversary model with malicious users, which can potentially cause a DoS (Denial-of-Service) or, in case of multiple malicious users, a DDoS (Distributed Denial-of-Service) attack. Also, there are no trust guarantees on the Backend concerning Confidentiality, Integrity, and Availability (CIA)~(Table~\ref{tab:cia-analysis}). 

\vspace{\baselineskip}
\subsubsection{DDoS on the Backend}
A device may intentionally or unintentionally cause a DoS on the server by sending multiple requests to notify a list of devices previously in range. Additionally, a DDoS can happen with multiple devices performing the same operation. Furthermore, the same concern should be considered to prevent one or multiple servers to flood one or multiple devices. Therefore, the notification schema should be carefully designed to prevent flooding of messages on both sides (backend/application). Furthermore, such an attack scenario is mitigated by a combination of two mechanisms: \textit{(i)} by adding an anonymous authentication when publishing data to the Backend, such as via a dedicated token only hospitals may use or by applying a proof-of-work activity and \textit{(ii)} by limiting the size of the message a single device can report to a sensible maximum value, such as that a device can only contact 1,000 users at most. 

\vspace{\baselineskip}
\subsubsection{Impersonification of the Backend}
Malicious users may try to impersonate the server in order to intercept the connection of multiple devices to the Backend. Thus, it is important to use a secure communication channel with the Backend to prevent malicious devices from acting as man-in-the-middle.
 
\vspace{\baselineskip}
\subsubsection{False Notification Reporting}
It is also important to prevent users from issuing false notifications, either maliciously or unintentionally. Thus, it is necessary to ensure that the application issues a “confirmation page/button” before changing a status and also (and not less critical) prevent multiple changes of status over a period of time. Although it had not yet been confirmed, but so far there are only very few cases of multiple COVID-19 infections on the same individual, thus, there does not seem to exist a reason to allow a device to change its status to “infected” or “not healthy” twice or more often. Thus, it is highly relevant to foresee such a warning/confirmation page/button. Furthermore, the Backend should prevent the situation, where multiple individual notifications are sent to a recipient device in multiple copies. Therefore, the mitigation measure will follow the basics in terms of DDoS mitigation as described above: the anonymous authentication when publishing data to the Backend prevents such falsified notification reports. 

\subsection{Privacy Enforcements}
\label{ssec:privacy_enforcements}
 
The enforcement of those privacy properties as set-up in Subsection~\ref{sec:privacy_properties} is key and addressed in the following way.

\vspace{\baselineskip}
\subsubsection{Privacy from Snoopers}
Since WeTrace will broadcast a signal on ``close contacts'' based on public keys generated (cf. Figure~\ref{fig:sequence_diagram}) so that others in close proximity can detect and possibly register such a contact anonymously, snoopers will also be able to see all public keys being advertised. However, since those public key-based identities will be valid only for a very limited time (those can be further shortened if needed), the user will not be more exposed to snooping than he/she already is with a Wi-Fi-enabled device that is broadcasting its MAC (Medium Access Control) address without any consent or knowledge of the user. This is still happening in case MAC randomization is enabled.

Also, only if the snooper was in real close contact, he/she will receive a notification of infection, but will not be able to track down from which user this notification was sent. That is due to the fact that for the notification it does not matter which information the snooper collects, but only which information the user’ devices collect and acknowledge as a close contact. Thus, this property is covered in an almost perfect manner by WeTrace.

\vspace{\baselineskip}
\subsubsection{Privacy from Contacts}
Privacy from contacts is addressed in a similar argument as for the case of ``Privacy from Snoopers''. Close contacts will receive a notification, if a user chooses to broadcast his/her infection status. However, all close contacts without exceptions will \textit{not} know from whom this message originated. A single case, where this could be inferred, would be where the user \textit{only} had close contact with a \textit{single} contact during the last 14 days and that contact would broadcast this information. While this is clearly not impossible to happen, its likelihood is very small, thus, WeTrace covers in the large majority of realistic cases this property, too. 

\vspace{\baselineskip}
\subsubsection{Privacy from Authorities}
Due to the fact that only encrypted messages are sent to the Backend (\cf Subsection~\ref{ssec:stakeholder_definitions}), it does not have access to any personal user data from any of these messages relayed. Thus, communications from or to the Backend does not reveal, \eg the number of infections or the identifiers of any recipient. The Backend only knows that someone wants to inform about a medical status, hence the existence of related communications only reveals that ``at least 1 medical status update had happened''. 

Additionally, if WeTrace would introduce random messages, any third party, such as an attacker or any authority, will not even know about such a medical status change. Therefore, this property is achieved by WeTrace as well.

\subsection{WeTrace Attack Vectors and Other Mitigations}

The basic WeTrace Architecture and its stakeholders involvement does potentially give raise to concerns, which had been identified as weaknesses or attack vectors, but are addressed also directly with suitable mitigation means. While two security-related aspects include possible attacks against the privacy requirement (a malicious scanning of advertisements sent of arbitrary devices and active message injections in combination with eavesdropping), the respective countermeasures are introduced below. A BLE-related protocol concern exists with respect to its limitations of broadcast messages, such that at worst public keys could not be broadcast, thus, a suitable operation is defined to circumvent this problem. And finally, performance concerns may raise with respect to the scalability of message decryptions, especially in case of many close encounters. 

\vspace{\baselineskip}
\subsubsection{Malicious Scanning of Advertisements}
A remaining privacy concern is due to the fact that a malicious user or attacked could start tracking a users' location by scanning his/her advertising packets. This can potentially happen, if the attacker is in the proximity of the user under attack. However, this case does not occur in reality, since within step 1 as above, the device's generation of a key pair is in fact the generation of a so called ``master seed''. This master seed is used to deterministically derive in turn an unlimited number of key pairs. This is designed in a way that the user will be changing the key in a specified period of time (for instance every \eg 30~min), making him/her traceable with that public key for that time frame only. Thus, the local knowledge of this device's validity period and the storage of the respectively applied key pair remain at the discretion of the user's device only and is fully decentralized. This leads to the situation that even a maliciously collecting Backend would possibly collect ``different'' public keys, which cannot be mapped onto a single device by any means. 

The major advantage of this approach is --- besides its elegance of hiding temporary identities even further very efficiently --- that the user still only stores one master seed and derives upon the reception of a notification from the Backend all asymmetric key pairs used during the past 14 days. Using those derived key pairs the device tries to decrypt the message by iterating over those keys. This does only require the storage of master seed only, since the dynamic derivation of all keys generated is time-wise not costly, but saves valuable storage. However, the implementation of the alternative, storing all keys generated over a 14~days period can improve the decryption time at the cost of a higher local key storage size.

\vspace{\baselineskip}
\subsubsection{Message Injections and Eavesdropping}
In the approach developed the risk of an attacker injecting packages and messages instead of eavesdropping exists. Thus the risk evaluation deals with the question, whether that is better or worse for a snooper. 

On one hand, selected tracing applications and proposals are today still prone to eavesdropping \ie PEPP-PT (Pan-European Privacy-Preserving Proximity Tracing). Eavesdropping refers to the fact that an attacker passively listens to all communications going on, which requires in case of BLE communications to be in close proximity or to install maliciously a BLE-based Backend, which collects all local communications to forward it to the attacker's infrastructure for analysis --- none can be prevented from happening. On the other hand, tracing applications and proposals are today still prone to message injections \ie WeTrace here. The major commonalities and differences are summarized as follows:

\begin{itemize}
    \item All unauthorized listening and eavesdropping is undetectable in the general case.
    \item Message injection is in general difficult or even impractical if these messages are ``directed at someone'' explicitly. 
\end{itemize}

Thus, only an approach, which can prevent message injection, while being prone against eavesdropping, can survive a security analysis. For WeTrace that means: In order to know, if User A was infected, the attacker would need to ensure that User A \textit{only} receives the message injected, since if multiple other users would also receive that message, the attacker would not be able to distinguish anymore from whom the message was received.

Furthermore, if the attacker relies on the fact that the ``reporting'' party needs to record his details, that party has control over ``when'' to record the attacker as a ``close contact'' --- that party can define, \eg how high the signal level needs to be or how long the contact needs to last. 

Finally, for the eavesdropper's scenario the attacker would just collect as much as possible and the other party has no control over what is being recorded or not. However, these data collected are of no use for the attacker, since the temporary identities change over time frequently and cannot be associated under any measure reliably with a device, thus, a user. 

\vspace{\baselineskip}
\subsubsection{Limitations of BLE Broadcast}
The BLE protocol is limited with respect to how many bytes can be broadcast while in background operation, thus, the major concern is, does public key fit into a BLE broadcast?

WeTrace requires for a secure encryption at least 24 Byte for the public key, ideally it would above 32 Byte. While the BLE advertisement message is limited with respect to the number of bytes being included as a payload, solutions exist to work around this limitation, even if the payload would be limited to only 16 Byte. The following options exist: 

\begin{itemize}
    \item Do not broadcast the entire public key. In this option only the first \textit{n} bytes (\textit{n} being the number of bytes to be included into the payload) are broadcast. Afterwards, the remaining 32-\textit{n} bytes are published to a server, for instance as a map that uses as the lookup key the hash of the first \textit{n} bytes. This will allow for the operation with an infinite size of keys, since only those devices that had been able to collect these \textit{n} bytes will be able to request the remaining bytes. Thus, the authority running the server will not know by any chance the full public key, but only 32-\textit{n} bytes.
    \item Use multiple advertisement packets with an  \textit{n} bytes payload. Since a possible contact counts as a ``close contact'' only after a certain amount of time, it is clear that a device has to collect at least 2 separate advertisement packets to be able to define that duration. Thus, it is viable for WeTrace to split up the key into multiple advertisement packets.
    \item Advertise only a fixed service Universally Unique Identifier (UUID). An UUID allow others, users and devices, to request the characteristics of that service. In this case, the characteristics are determined by those 32 Byte.
\end{itemize}

\vspace{\baselineskip}
\subsubsection{Scalability of Message Decryption}
The WeTrace approach and protocol outlined requires a user to decrypt all messages to understand (being able to correctly interpret the content of an initially encrypted message), if a message was directed at him/her. While this scales reasonably well as long as the user has to decrypt up to 1 million messages, it might become a problem with larger numbers, especially because the drainage of the users' device battery with decryption tasks needs to be avoided. However, under the assumption of COVID-19 curfews as well as lockdowns, the likelihood that mobile devices are recharged more often is large due to having fixed power supplies. 

A suitable path to mitigate this performance aspect without the considerations of more frequent recharging options is by prefixing the message with the first \textit{n} bytes of the hash of the public key. This would allow the user to select at first and reduce drastically at second the number of messages he/she needs to decrypt. Such an approach will basically allow a device to cope with almost any reasonable number of encrypted messages in those cases of hundreds, now even thousands of close encounters.


%% file: sections/observations.tex
\section{Observations and Analysis}
\label{sec:observations}

The WeTrace approach and application --- as it had been designed --- suits its needs. The details of these are first discussed and secondly evaluated from the perspective of advantages over other related work and drawbacks compared to related work --- including already their mitigation means --- as of below. While major conclusions are drawn, too, it is essential to observe, which trade-offs have been taken into account and how next steps for tracking and tracing applications in case of a pandemic are foreseen. 

\subsection{Trade-offs}
Unfortunately, there are many trade-offs to consider. In the theoretical view of the design WeTrace follows, it trades off \textit{(a)} a central analysis versus \textit{(b)} the privacy of users and user data. While it is obvious that a central analysis of data can be advantageous, once authorities for instance want \textit{(i)} to detect ``hot spots'' of infections or \textit{(ii)} to perform page ranks on possible next infections, such data being processed will have to be stored either centrally or locally, while for the latter access to authorities have to be guaranteed, too. Thus, authorities in case of \textit{(a)} will know much more on the participating users' behavior than necessary. 

In essence, this additional information is for the successful analysis, prediction, and action plan development of COVID-19 cases not needed, since the measure of ``proximity'' is based on the evaluation of epidemiological requirements fully sufficient. As noted above, on an individual basis and designed as an opt-in approach, individuals may add location information and time, too. In case of large cities, this is very unlikely to impact the user's privacy, in case of rural locations, where only a few dozens of inhabitants reside, such decisions may be considered to be more critical with respect to the privacy aspect, however, it was clearly stated to be an option, thus, freely made available data does not violate privacy regulations. 

\subsection{Proximity Discussion}
Furthermore, a key requirement in WeTrace is that the application has to see both devices needing to record one-another, the proximity. This is only and solely based on the use of public-private cryptography, for which such key pairs may be generated on the spot, since there is no need to register these key pairs at a Certification Authority due to the fact that not the individual's identity is the key, but the fact that two individuals exchange the proximity information at first and may exchange later infection status without revealing any identity for that second step, only the public key once collected at the close encounter. Thus, this approach enables WeTrace to encrypt the message for a possible receiver or receivers of the encrypted report and does fulfill the trade-offs as of \textit{(b)}. A symmetric encryption scheme, obliviously, will not work, and the full independence of any centralized authorities does establish an exchange model of status information on an ad-hoc basis without any centralized control. 

The proximity requirement is met, because WeTrace defines a close contact as an individual being in a distance of 2~m of proximity to a COVID-19 infected person and for an epidemiologically relevant period of 10 to 15 minutes. These parameters are in accordance with several standards defined by major health organizations worldwide. For example, the US Centers for Disease Control and Prevention (CDC) defined ``close contact'' as ``\textit{...being within approximately 6 feet (2 metres) of a COVID-19 case for a prolonged period of time}''~\cite{cdcUSCloseContact}, the European CDC defines as ``\textit{...having had face-to-face contact with a COVID-19 case within 2 metres and more than 15 minutes}''~\cite{ecdcCloseContact}, the New South Wales Ministry of Health defines as ``\textit{...greater than 15 minutes face-to-face contact in any setting with a confirmed case in the period extending from 24 hours before the onset of symptoms in the confirmed case..}.''~\cite{nswGovCloseContact}, and the Brazilian Ministry of Health defines as ``\textit{A person who has had face-to-face contact for 15 minutes or more and at a distance of less than 2 meters}''~\cite{brCloseContact}. Thus, WeTrace's major parameters are aligned with major guidelines of the ``close contact'' definition worldwide.

\subsection{Privacy Discussion}

Concerning the privacy requirements listed in Section~\ref{sec:privacy_properties}, the WeTrace approach and application addresses all the imposed privacy requirements and challenges highlighted in~\cite{cho2020contact}. Hence, the approach tackles the following privacy aspects:

\begin{itemize}
    \item \textbf{Privacy from Snoopers}: WeTrace addresses this challenge by limiting the time-wise validity of public and private key pairs, generating a new one every $X$~minutes. Furthermore, the \textit{snooper} is not aware of the device's exact location nor its identification, the snooper is only aware of the notification that a close contact with an unknown infected individual took place.
    \item \textbf{Privacy from Contacts}: This property is tackled by encrypting the message with the public key from ``close contact'' devices. Thus, an individual will know that ``infection'' messages were sent to him/her, but not who sent them, since the Backend and the application does not store any private information.
    \item \textbf{Privacy from Authorities}: Similarly, with the employment of public-key encryption, all messages exchanged between the Backend and all devices are encrypted. They are only decrypted with the knowledge of the private key, which remains solely in the user's device. Thus, authorities cannot have access to the messages' content.
    \item \textbf{Infrastructure Requirements}: WeTrace requires a single Backend with a straightforward message broadcasting application. Even though a logically single server is required, multiple instances of such Backend can exist to increase their availability.
\end{itemize}

\subsection{User Overhead Discussion}

The WeTrace application resides as of today in a separate prototypical implementation, thus, the question on how to integrate this important privacy-preserving functionality into tracking and tracing apps, which focus on those layers needs to be answered. 

For example, the WHO's app (World Health Organization) as an ``e-Library of Evidence for Nutrition Actions (eLENA)''~\cite{eLENA} could place one example for such an integration. The WHO Zika App~\cite{ZIKA} as of Google Play can serve as a second one, once it is turned into a COVID-19 app. On one hand, any integration would need to carefully consider technical constraints, on the other hand, only applications, which do not require user credentials are suitable, since otherwise privacy may suffer and could be at risk. Thus, the WeTrace application will have to be offered as an SDK (Software Development Kit), which makes it much easier for any other application to integrate WeTrace. 

\subsection{Major Observations}

Overall, the tradeoffs and discussions highlighted indicate the key aspects of a system in which very many individual participants act as in one role (inhabitants) and only very few act in the second role (authority). Thus, the WeTrace design decision taken does enable the two roles to act as they are required to act, independently, however based on each other, and the method implemented shows all properties in which the privacy requirement of the proposed solution is integrated very elegantly and easy to implement. 

Of course, the important next step will be that the ``community'' of major players and stakeholders can agrees on a ``standard'' on how to trace infections in case of COVID-19 and then to ensure that all developers will use the same standard or protocol such that the system can profit from a network effect across different applications, regions, and even countries. It is imperative that this initiative, to which WeTrace as well as many other applications need to be counted to, is following a open source philosophy, so that \textit{(i)} security-related measures can be verified openly, \textit{(ii)} functionality verification can be performed at no risk, \textit{(iii)} various application developers can cooperate, and \textit{(iv)} all stakeholders involved can collect those data, which are essential and securely collectable. 

\subsection{Evaluations}
\label{sec:evaluations}

The investigation of related work in tracking and tracing applications on the case of COVID-19 did reveal that the problem is not the collection of data as such, typically provided by accessing mobile devices such as smartphones, which are in the possession of an individual, but the guarantee that those data collected are fully maintaining the basis and all of the relevant details of a privacy-protected approach. Thus, the human individual and his/her privacy, his/her private data, and a fully anonymous processing of related data is the key to meet European and many other countries demands, while at the same time being compliant especially with the European regulation, especially GDPR. 

\subsubsection{Communication Channel Evaluation}
Thus, the WeTrace approach as described above utilizes the Bluetooth communication channel, which many modern mobile devices provide today. This coincides with the low range requirements of the medical dimension, since infections potentially can only happen in case of close proximity where humans need to stay within below 2~m for approximately 10 to 15 minutes. 

The pure knowledge of such proximity determines the most essential information for epidemiologists, since based on density-related information, not necessarily including the exact geographical location, but a region only, prediction models of spreading rates or relaxing measures can be derived. However, since proximity and location determine a highly valuable good for every single person and individual on earth, it need to be fully protected from possible misuse or unintended use. Just imagine the value of a human's geographical position for marketing, commercial services, or monitoring? This threat for an open society has to be balanced with the medical and health threats COVID-19 imposes on the society. WeTrace allows for both to be reached and maintained at a highly secured level of operation. 

Furthermore, the WeTrace application requires between 24 and 32~Byte to be transmitted via the BLE communication channel. Unfortunately, that is technically limited in case of iOS-based devices, since two advertisement packets are required here. However, firstly, this ``loss'' of a single packet approach is not crucial to the game, since WeTrace requires the reception of two advertisement packages always to measure the time a possible ``close contact'' had taken. Thus, the potential limitation to technically one message only does not harm at all. Secondly, the very strong privacy-preserving approach adds dedicated overhead to the approach, since a human associated with a smartphone can consider himself/herself ``infected'' or ``uninfected'' only, once all relevant data had been decrypted, which causes potentially in the general case a higher compute burden. However, this drawback can be mitigated already by adding the first few bits of the relevant public key into the message being communicated, such that only those messages need to be decrypted, which provide a partial match to the owner's public key.

\subsubsection{Privacy and Attack Evaluations}
The application of the well-known asymmetric cryptography only allows for the deciphering of a message at that destination it had been intended for, since that human operating that mobile device may remember his/her private part of the keys. And all proximity-related messages are sent in an encrypted manner over BT in a low range setting. Since additionally literally every other potential participant only listens to random data, even a brute force attack will not succeed to decrypt messages reliably on the fly. Therefore, WeTrace is the only known approach so far, which ensures that any receiver of a message knows that this is for him/her, but does not know who the original sender was.

Furthermore, the users deploying WeTrace are offered an opt-in path to decide, whether they want to add to the proximity message additional information, such as the exact location (not only a region) and the time. Thus, the application does operate in an open source manner only on the very basic and privacy-protected data needed to crowd-source data to help COVID-19 countermeasures to be based on currently measured details. 

While this is considered to be a clear advantage, even further relevant attacks are mitigated. A passive collection of communications in such a certain physical near range will not provide any reasonable amount of information, which could be used to reveal the sender's identity. Although as outlined above, potentially the injection of public keys is possible in any setting, it can only happen if and only if the attacker is ``local'' for a certain amount of time. Thus, the WeTrace approach developed does not suffer from this attack, since it alone, the application, does configure and decide on the Received Signal Strength Indicator (RSSI) and the time. Therefore, eavesdropping does not show any negative impacts. 

Finally, the WeTrace application addresses all challenges that have been highlighted in~\cite{cho2020contact}. This means that the WeTrace application complies with all imposed privacy requirements and especially with the respective general demand and key detailed requirements.

\subsubsection{Data Storage and Size-related Evaluations}
Since the Backend just stores all encrypted messages of the last 14~days, it offers these to whoever requests them. Overall, this is static and random data which needs storage space. 

Concerning especially the data stored on the client side (\ie the smartphone) WeTrace safely assumes an individual encounters with 5,000 other smartphones in 14~days as close contacts (\ie $\le$ 2~m every 4~min) and changing the key every 15~min. Furthermore, assuming a 4~Byte size for longitude, 4~Byte for latitude, and a 4~Byte length for the timestamp, an encounter message will have the size of 12~Byte. Considering that 4 new keys are generated every hour for 24~hours and 14~days, an individual will own 1.344 private keys generated and 5.000 encounter messages stored. Thus, the device, will need to decrypt 6.720.000 messages (\ie approximately equaling 80~MByte of data considering the 12~Byte-sized message). For today’s phones this is not considered to be a large number, due to an average of 80~GByte of storage capacity per smartphone~\cite{smartphoneStorageSize}. It should be noted as well that this is already an extreme example and it is unlikely to happen in daily operations. Thus, the client's side performance is not a critical aspect within the WeTrace application.

\subsubsection{Scalability Evaluations}
The overall scalability of WeTrace depends on set of factors: \textit{(a)} the number of infections, \textit{(b)} the number of close contacts, and \textit{(c)} the number of keys. Thus, if these numbers grow also the product grows. Currently, a smartphone is able to decrypt approximately 1 million messages within seconds, which is acceptable. However, if these numbers grow in the scale of billions, the scalability has to be mitigated: every message is prefixed with \textit{n} bits of the public key. By doing this, the device will only try to decrypt those messages, which match with the first \textit{n} bits of their public key. This straightforward and easy to implement scaling strategy allows for an exponentially cut down of the number of messages to be decrypted. \textit{I.e.,} if a 1~bit prefix is assumed, the reduction of decryption sets is at 50\text{\%}, with a 2~bit prefix 75\text{\%} are achieved. However, the prefix should remain at an overall size, where the number of bits being disclosed does not reveal too much about the actual public key. 


%% file: sections/conclusions.tex
\section{Summary and Conclusions}
\label{sec:conclusions}

The protection of people and society against harm and health threats involves a variety of different disciplines. While in case of the COVID-19 pandemic, the virus and its medical treatment -- from currently effected patients to the vaccination of future people -- do see a major focus of research and work, the data collection of very basic and health-related data of individuals in today's highly mobile society does help to plan, protect, and identify next steps health authorities and governments can, shall, or need to plan for or even implement. Thus, every individual, every human, and every inhabitant of the world is the key player -- very different to many past crises'. 

Although the involvement of all humans cannot be considered to be negative as such, the individual's \textit{(a)} health and \textit{(b)}  privacy shall be considered in a very carefully crafted balance, not overruling one with another or prioritizing one aspect. If the solution of the current pandemic's data collection can be based on a fully privacy-preserving application, which can be used by individuals on their mobile devices, such as smartphones, while maintaining at the same time their privacy and while respective data collected in such a fully distributed setting does help to confine the pandemic, an important step forward can be achieved in a democratic and very open, but still and especially privacy-protecting world.

Thus, the WeTrace approach as described in this paper utilizes the Bluetooth communication channel, many modern mobile devices provide in a way, where asymmetric cryptography being applied only allows for the deciphering of a message for that destination it had been intended for. Since literally every other potential participant only listens to random data, even a brute force attack will not succeed. WeTrace is the only known approach so far, which ensures that any receiver of a message knows that this is for him/her, but does not know who the original sender was. 

Besides this clear advantage, even a passive collection of communications in a certain physical range will not provide any reasonable amount of information, which could be used to reveal the sender's identity. Although potentially, the injection of public keys may be possible, if and only if the attacker is ``local'' for a certain amount of time, the approach developed does not suffer from this attack, since it alone does configure and decide on the Received Signal Strength Indicator (RRSI) and the time. Therefore, eavesdropping does not show any negative impacts. 

Finally, as a slight drawback of this very strong privacy-preserving approach is only the overhead to determine, if a human associated with a smartphone can consider himself/herself ``infected'', since all relevant data needs to be decrypted. However, this can be mitigated already by adding the first few bits of the relevant public key into a message communicated, such that only those messages need to be decrypted, which provide a match to the owner's public key.

\vspace{\baselineskip}

In conclusion, the WeTrace application provided in close relation to those requirements being defined and evaluated a highly suitable system based on the BTE communication channel in support of crowd-sourcing for COVID-19-relevant data in a privacy-protecting setting. This approach is scalable as well since close proximity of humans can be considered in the range of a few hundreds of people, not thousands anymore, since these are legally forbidden. Therefore, in case a mobile device would see way too many messages, a possible alarm can be raised, which by itself already identifies that a violation of meeting regulations had occurred. 

In the same line of arguments, the resource consumption of mobile devices is not at stake, since especially data to be stored is limited to the public keys of those messages received. While the overhead on the compute side had already been mentioned, and it is considered to be at the lower end of the spectrum, the legal compliance with especially privacy considerations of users and humans have been met in full. 

Note that full-fledged performance evaluation of this approach and the WeTrace application has not been performed. However, the open-source implementation is available at~\cite{WeTrace} and related soft-requirements' suitability of thresholds not reached in practice have been discussed above.